\documentclass[useAMS,usenatbib]{mn2e}
\usepackage{graphicx}

\usepackage{indentfirst}
\usepackage{afterpage}
\usepackage{subfigure}
\usepackage{ulem}

\title{Global Gravitationally-Organized Spiral Waves and the Structure of NGC 5247}
\author[Khoperskov et al.]
  {S.A.~Khoperskov,$^1$ A.V.~Khoperskov,$^2$ I.S.~Khrykin,$^3$
  V.I.~Korchagin,$^4$
  \newauthor 
   {D.I.~Casetti-Dinescu,$^5$ T.~Girard,$^5$
  W.~van Altena,$^5$ D. Maitra$^6$} \\
  $^1$Institute of Astronomy, Russian Academy of Sciences, Pyatnitskaya st., 48, 119017, Moscow, Russia \\
  $^2$Volgograd State University, Yniversitetskiy pr., 100, 400062, Volgograd, Russia\\
  $^3$Physics Department, Southern Federal University, Zorge st.,5, 344090, Rostov-on-Don, Russia  \\
  $^4$Institute of Physics, Southern Federal University,
      Stachki st. 124, 344090, Rostov-on-Don, Russia\\
  $^5$Astronomy Department, Yale University,
      New Haven, CT 06520-8101, USA\\
   $^6$ Department of Astronomy, Univ. of Michigan
500 Church St., Ann Arbor, MI 48109-1042, USA}

\date{Released 2012 Xxxxx XX}

\def\LaTeX{L\kern-.36em\raise.3ex\hbox{a}\kern-.15em
    T\kern-.1667em\lower.7ex\hbox{E}\kern-.125emX}

\newcommand{\kmps}{km s\ensuremath{^{-1} }}	
\newcommand{\mybf}{\bf}

\begin{document}

\label{firstpage}

\maketitle

\begin{abstract}

Using observational data, we build numerical N-body, hydrodynamical and
combined equilibrium models for the spiral galaxy NGC 5247. The
models turn out to be unstable towards spiral structure
formation. We simulate scenarios of spiral structure formation for
different sets of equilibrium rotation curves, radial velocity dispersion
profiles and disk thickness and demonstrate that in all cases a
simulated spiral pattern qualitatively agrees with the observed morphology
of NGC 5247. We also demonstrate that an admixture of a gaseous component
with mass of about a few percent of the total mass of the disk increases a
lifetime of a spiral pattern by approximately 30\%. The simulated
spiral pattern in this case lasts for about 3 Gyr from the beginning
of the growth of perturbations.
\end{abstract}


\section{Introduction}
Attempts to understand the phenomenon of spiral structure in galaxies
have a long history. It has become clear nowadays that the
spiral structure is a density wave (or waves) propagating in a
multi-component stellar-gaseous disk. A universal mechanism for
the generation of such spiral density waves that successfully explains
the rich morphological variety of spiral galaxies does not yet exist.
To describe the observed patterns, a few spiral-generation mechanisms
are usually invoked.  Some researchers treat the spiral structure as
long-lived global modes that last in the galactic disks for tens of
galactic rotations \citep{Bertin1989a,Bertin1989b,Bertin1996}. Others consider the
spiral structure as a transient phenomenon, so that the spiral
pattern changes many times during the galactic evolution \citep[see,
e.g.,][for references]{Sellwood2011}.
\citet{Gerola1978} even suggested that spiral structure is merely
a product of recent star formation, outlined by the new-born stars.
On observational ground, the problem of the origin of spiral
structure was raised recently by~\citet{Eskridge2002,Elmegreen2011, Kendall2011}. These authors found
that galaxies which are optically grand design spirals, show
a grand design structure in the near-infrared as well. The spiral
structure therefore is not merely a manifestation of a recent
star formation but represents also a spiral structure in the underlying
stellar mass distribution. The paper by~\citet{Elmegreen2011} shows that some flocculent spirals remain flocculent even in the IR. This observational evidence rules
out at least one approach which suggested that the spiral structure in
galaxies is a pattern of a recent star formation and does not involve
an underlying stellar mass distribution.   To understand to what extent
the various suggested explanations are consistent with the observed
phenomena of spiral structures, a detailed, quantitative
comparison between the theoretical predictions and the observed
properties of the spiral patterns in galaxies should be undertaken. In
this paper, we continue our efforts to compare the properties of the
observed spiral structure with the theoretical predictions. In previous
papers \citep{Korchagin2000,Korchagin2005}, we made such a comparison
using a hydrodynamical approach to model the dynamics of galactic disks.
In this paper, we simulate three-dimensional collisionless,
and collisionless-gaseous models of galactic disks, aiming to determine
the model parameters that most closely match the observed spiral
pattern in the galaxy NGC 5247. We choose this galaxy for our comparison
because it is well studied observationally, and its rotation
curve as well as disk velocity dispersion and luminosity distribution have been accurately
determined.

In the cases that have been studied, the admixture of
a gaseous component increases the lifetime of a spiral pattern
\citep{SemelinCombes2002,Fux1999}. In general however, the lifetime of
the spiral patterns in isolated galaxies and the role of gas remain an
open issue, that we address in this paper.

In Section 2 we discuss the observational properties of the galactic disk
of NGC~5247. In Section 3 we present the basic equations and discuss the
disk equilibrium. Section 4 presents the results of numerical simulations
of a gaseous and a collisionless stellar model of the galaxy NGC
5247. Section 5 presents the results of our simulations of stellar-gaseous
models, and in section 6 we present the concluding remarks.

\section{NGC 5247: constraints from observational data}

\begin{figure}
 \includegraphics[width=1.0\hsize]{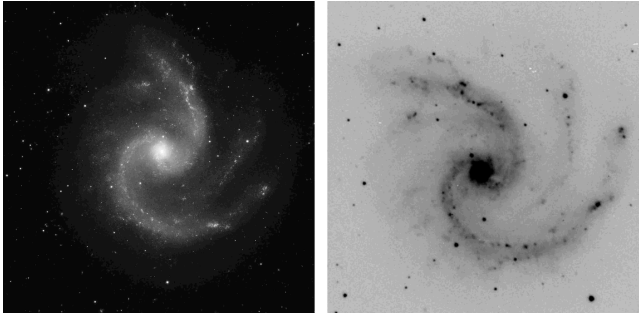}
  \caption{Left: optical image of the NGC 5247 galaxy (\citet{Eskridge2002}). Right:
  HII regions outlining the spiral arms.}
 \label{fig::obs}
\end{figure}

NGC 5247 shown in Figure~\ref{fig::obs} is a nearby spiral galaxy  which probably belongs to the Virgo Supercluster of galaxies  (Bingelli et a. 1985). It has a strong, well developed two-armed
``grand-design'' spiral pattern and does not demonstrate any signs of
interactions with other galaxies \citep{Considere1988}. The spiral pattern however looks somewhat lopsided. A low-amplitude spiral pattern is detected in an interarm region. However
the morphology of the galaxy is dominated by the two-armed spiral. The photometric data do not point to any noticeable bar in the central regions of the disk, therefore NGC 5247 is a classical example of a grand-design spiral galaxy.

Our numerical modeling is based on photometric data taken by
\citet{Zhao2006}.  We take as the radial scale length of the disk $r_d = 4.8$ kpc as determined by these authors.
The inclination angle of NGC
5247 is rather uncertain. Measurements range from $i=20^\circ$
\citep{Zhao2006} to $i=40^\circ$ \citep{HLeada}.  In this paper we adopt an inclination of $28^\circ$ which is close to an average from these references, and
a distance to the galaxy of $17.4$~Mpc \citep{HLeada}

The vertical scale of the disk ($z_0$) significantly influences
its stability properties. Unfortunately, an accurate estimate of
the vertical thickness of the galactic disk is not simple.
\citet{Zhao2006} tackled this problem by using the Jeans equations
applied along the vertical axis of the disk, and estimated the disk
thickness of NGC 5247 to be $z_0=1.5\pm 0.6$ kpc. With this value, the ratio
of the radial to vertical scale length is $r_d/z_0 = 3.2^{+2.1}_{-0.9}$.

\begin{figure}
\includegraphics[width=1.0\hsize]{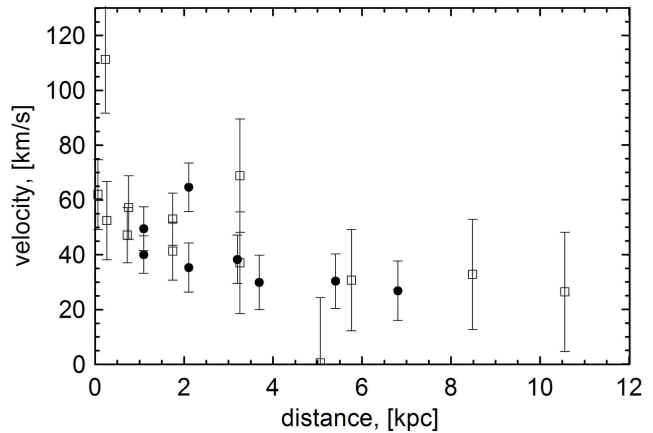}
\caption{Radial dependence of the observed velocity dispersion of
the stellar disk of NGC 5247. Filled circles --- data from \citet{Kruit1986}, open squares --- measurements of \citet{Bottema1993}.}
\label{fig::ObsCz}
\end{figure}

The line-of-sight velocity dispersion as a function of
galactocentric radius for NGC~5247 has also been determined from
observations \citep[][see Fig.~\ref{fig::ObsCz}]{Bottema1993,Kruit1986}.
The measurements by both groups agree with each other and show
approximately exponential decrease of the velocity dispersion with radius,
which is typical for disk galaxies.

\section{{\mybf Modeling NGC 5247:} Basic equations}

We assume that the potentials of the dark halo and the stellar bulge
are static and remain axisymmetric. Such an assumption does not affect the dynamics of a galactic disk because
a contribution of high velocity dispersion components to the spiral density waves is insignificant even if they involve most
of the galactic mass \citep{Marochnik1972}.
The galaxy does not have a large central bar, and we exclude from a consideration the models with a bar which
is considered to be an effective generator of the spiral density waves
(\citet{KorchaginShevelev1981,Buta2009,Kaufmann1996}). Second, observational data do not allow us to estimate the degree of asymmetry of spheroidal subsystems, and the emergence of new free parameters greatly complicates the analysis. Moreover, an interaction with a live bulge or a halo can not be decisive.

To simulate the dynamics, we use a set of equations which
self-consistently describe the behavior of a stellar-gaseous disk in
equilibrium with an external gravitational field $\Psi_{ext}$.  The
set of equations for the stellar disk consisting of $N$ gravitationally
interacting particles is written as:
\begin{equation}\label{eq::nbt}
 \frac{d^2 {\bf r}_i}{d t^2} = -\nabla\big[ \Psi_s({\bf r}_i) + \Psi_{ext}({\bf r}_i) + \Psi_g({\bf r}_i) \big]\,,
 \ \ i=1,...,N\,.
\end{equation}
Here $\Psi_s$ is the gravitational potential of the stellar
component of the disk, $\Psi_g$ is the potential caused by the gaseous
component, and $\Psi_{ext}$ is the combined gravitational
potential of the galactic halo ($\Psi_h$) and the bulge ($\Psi_b$).
Therefore $\Psi_{ext} = \Psi_h + \Psi_b$.

The set of equations describing the dynamics of the gaseous
component are given by:
\begin{equation}\label{eq::hyd1}
\frac{\partial \rho_g}{\partial t} + {\bf\nabla \cdotp}
(\rho_g {\bf u}) = 0\,,
\end{equation}
\begin{equation}\label{eq::hyd2}
 \frac{\partial \rho_g {\bf u} }{\partial t} + {\bf u}\cdotp {\bf\nabla} (\rho_g {\bf u} ) + {\bf \nabla} p + \rho_g\nabla\big[
  \Psi_g + \Psi_{ext} + \Psi_s\big] = 0\,,
\end{equation}
\begin{equation}\label{eq::hyd3}
\frac{\partial E}{\partial t} + {\bf \nabla}
\cdotp [(E + p){\bf u}] + \rho_g{\bf{u}}\nabla\big[
  \Psi_g + \Psi_{ext} + \Psi_s\big]=0\,,
\end{equation}
here  $\rho_g$, $p$, and ${\bf u}$ are the volume density, the
pressure and the velocity vector of gas.
The energy density of gas
$E$ is given by the expression $E = \rho_g|{\bf{u}}|^2/2 + p/(\gamma
-1)$. We use an adiabatic equation of state $\rho_g =
K_s p^{\gamma}$ for the gas, where the value of the adiabatic
index is assumed to be $\gamma=1.1$ and the constant $K_s$ was
chosen so that the sound speed is equal to $c_s$=10 \kmps\ in central regions of the disk. The sound speed of
the gas decreases slowly with radius, reaching approximately 8 \kmps\ at the
periphery of the disk. Self-gravity is taken into account by the Poisson
equation applied for the gaseous and stellar components:
\begin{center}
\begin{equation}\label{eq::poiss_gas}
\Delta \Psi_g = 4\pi G \rho_g\,,
\end{equation}
\end{center}
\begin{equation}\label{eq::poiss_stellar}
\Delta \Psi_s = 4\pi G \rho_s\,.
\end{equation}

To model the gravity of the halo, we use the potential:
\begin{equation}\label{eq-halo}
    \Psi_h= \frac{G M_h}{C_h} \cdot \left\{
\ln(\xi) + \frac{{\rm arctg(\xi)}}{\xi} + \frac{1}{2}
\ln\frac{1+\xi^2}{\xi^2}  \right\} \,.
\end{equation}
Here $\xi = {r}/{a_h}$, $a_h$ is the scale length of the halo potential,
$M_h$ is the mass of the halo within radius $r_h=20$ kpc and $ C_h
= a_h (r_h/a_h - {\rm arctg}(r_h/a_h))$. Such a choice provides a constant
rotation velocity at large radii $r > 2a_h$, so we have a rotation
curve typical for disk galaxies.

The King model with a cutoff a density distribution of large radii
$r > r^m_b$ is used to model the potential of a stellar bulge:
\begin{equation}\label{eq-bulge}
 \Psi_{b} = - \frac{ G M_b }{ C_b r  } \ln \left[ \frac{r}{r_b} + \sqrt{1+\left(\frac{r}{r_b}\right)^2} \right],
\end{equation}
here $r_b$ is the bulge scale length, $M_b$ is the bulge mass within $r_b^{m}$ and $ \displaystyle C_b = \ln\left(\frac{r_b^{m}}{r_b} + \sqrt{1+\left(\frac{r_b^{m}}{r_b}\right)^2}\right) - \frac{r_b^{m}/r_b}{\sqrt{1+(r_b^{m}/r_b)^2}}$.

The observed rotation curve of the disk can be reproduced by varying the
parameters of the dark halo ($M_h$, $a_h$), the stellar bulge ($M_b$,
$r_b$) and the stellar disk ($M_s$, $r_d$, $z_0$).

To analyze a behavior of a one-component collisionless model, we assume that the gas density $\rho_g$ is equal to zero, and solve equation (\ref{eq::poiss_stellar}) using TREE-code \citep{Barnes1986} adapted to the parallel calculations. The number
of particles used in our collisionless models range from
$N=10^6$ to $10^7$. To build a self-consistent model of a stellar-gaseous
disk, the set of hydrodynamical equations for a gaseous disk rotating
in a self-consistent gravitational potential is added to the equations
of motion for a stellar disk.

We apply a TVD MUSCL type scheme in the cylindrical coordinate system \citep{vanLeer1979} to solve the equations (\ref{eq::hyd1})--(\ref{eq::hyd3}).
To preserve conservation laws at a grid level, a finite-volume
approximation of the variables is used:
$$
(q)_{i+1/2} = q_i + 0.25((k+1)D_{+} + (k-1)D_{-}))_i\,,
$$
here $(q)_{i}$,$(q)_{i+1/2}$ are the conserved variables at
a border of $i$-th cell and at its center correspondingly; $D_{+}
= minmod(d_+,b d_{-})$, $D_{-} = minmod(d_-,b d_{+})$, where $d_-$,
$d_+$ are the variations of the conserved variables to the
left and to the right of a cell border. The parameter $b =
{(3-k)}/{(1-k)}$ determines an order of approximation in space.
We use the value $k=1/3$, corresponding to the third order of  spatial
approximation. The function
$$
minmod(x,y) =
\frac{sign(x)+sign(y)}{2}min(|x|,|y|)
$$
 is used to reconstruct the
discontinuous functions between the nodes of the computational
grids.

Interaction between the stellar and the gaseous components of the disk
occurs due to gravity. Similar to the one-component models, such
interactions are computed using the TREE-code.

At the beginning of the simulations, the stellar disk
is in an equilibrium in the radial and the vertical directions with its density given by:
$$
\rho_s = \rho_{s0} \exp(-r/r_d)\cdot A(z/z_0) \,,
$$
here $r_d$, $z_0$ are radial and vertical scales of the stellar
disk. Function $A(z)$ determines a vertical distribution of density of the stellar disk: $A(z) = {\rm ch^{-2}}(z/z_0)$.

The equilibrium of the stellar disk in the vertical direction is
determined by solving the Jeans equation:
\begin{eqnarray}\label{4--Eq-VelocityRotation-eqJEANS}
\rho_s\,\frac{d}{dz}\left( c_z^2 \frac{d\rho_s}{ dz}\right) -
c_z^2 \left( \frac{d\rho_s}{ dz} \right)^2 + \nonumber \\
 + \rho_s^2\,\frac{d^2c_z^2}{dz^2} + 4\pi G \rho_s^2 \big( \rho_s
+ E+\rho_{ext}(z)\big) +
\rho_s^2\,\frac{d}{dz}\frac{E_\alpha}{\rho_s} = 0 \,,
\end{eqnarray}
 $$
 E=-\frac{1}{ 4\pi G r}\frac{\partial V^2_{c}}{ \partial r}\,, \ \ \ \
 E_\alpha = \frac{\partial (r\rho_s\alpha_{rz})}{r\,\partial r}
 \,.
 $$
Here $c_z$ is the stellar velocity dispersion in the vertical
direction, $\alpha_{rz}=\langle u w\rangle$ is the result of
averaging the product of the radial $u$ and vertical $w$ velocity components, $\rho_{ext}$ is the total density of the halo and the bulge.

By using Jeans equation in the radial direction, we determine the
rotational velocity of the stellar disk in the disc plane $z=0$~\citep{Khoperskov2010}
\begin{eqnarray}\label{4--Eq-VelocityRotation-eqJEANS}
V^2=(< v >)^2 = V_c^2 + c_r^2\, \Big\{ 1 - \frac{c_\varphi^2}{
c_r^2} + \nonumber \\
 +\frac{r}{ \rho_s c_r^2}\frac{ \partial (\rho_s c_r^2)
}{
\partial  r} + \frac{r}{ c_r^2}\frac{\partial  \alpha_{rz} }{ \partial  z}
\Big\} \,,
\end{eqnarray}
here $V_c$ is the circular velocity of a test particle in an
axisymmetric field $\Psi$:
$V_c^2/r = - \frac{\partial \Psi}{\partial r}$\,,
$c_r/c_{\varphi} = {2 \Omega}/{\kappa}$, $\Omega = V/r$ and
$\kappa = \sqrt{4\Omega^2\left( 1 + {r}\,{d\Omega}/({2\Omega}{d
r})\right)}$ is the epicyclic frequency.

 The gaseous disk extends in our models beyond the optical
radius of the stellar disk to the distance $r = 16r_d$.
The density distribution of the gaseous disk is modeled by the
expression
$$
\rho_g = \rho_{g0} (1 - r/(16r_d)) \cdot B(z)
$$
where the function $B(z)$ is determined by the balance of the gravity
force and the gas pressure gradient. With such a distribution, the gas
density decreases by a few times within the optical radius, and goes to
zero in the ghost zones of the computational grid at $r =16r_d$. The Euler equation uniquely determines the rotational speed of the gaseous disk:

\begin{equation}
 -\frac{v_{\varphi}^2}{r} = -\frac{1}{\rho_g} \cdot \frac{\partial p}{\partial r} -
 \frac{\partial }{\partial r}\left[\Psi_g + \Psi_s + \Psi_{ext}\right]\,.
\end{equation}

We build a set of models for a stellar-, and a stellar-gaseous
disk of the galaxy NGC~5247 such that the model predicted
quantities like rotation curve and  velocity dispersion  are
consistent with the observational data (and the uncertainties associated
with the observations).
In the following sections, we explore the model parameter space and
discuss the dependence of the
morphology of spiral pattern on the rotation curve, the stellar
velocity dispersion, the vertical thickness of a stellar disk, and
mass ratio of gaseous to stellar
component of the galactic disk.

\section{Disk equilibrium}

\subsection{Rotation curve}

\begin{figure}
{\includegraphics[width=1.0\hsize]{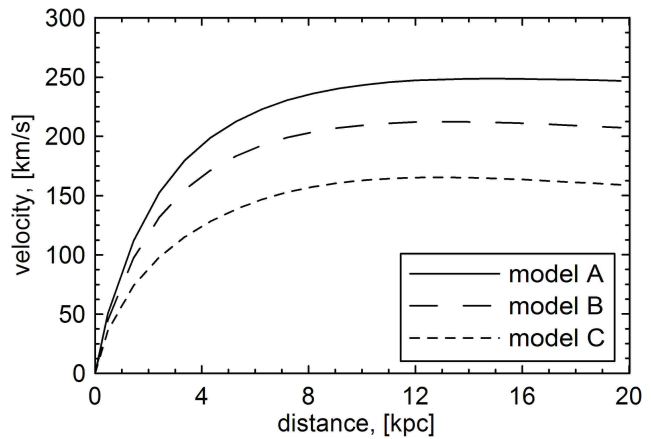}}
\caption{Rotation curves for one-component models.
The maximum values of the rotation curves are equal to 245~km/sec
(Model A), 205~km/sec (Model B) and 153~km/sec (Model C).
\label{fig::diff_rc}}
\end{figure}

 A small inclination
angle of the galaxy influences significantly a reconstruction of its
rotation curve. Using the Tully-Fisher relation, \citet{Patsis1997}
found the maximum rotational velocity of the galaxy to be:
$V_{\max} = 205$ km/s. Other measurements give the values $V_{\max}
= 165 \pm 20$ km/s \citep{Bottema1993} and $V_{\max} = 300$ km/s. Following the paper of \citet{Contopoulos1986}, we
build a set of rotation curves that have a quasi-solid rotation in the
central regions of the disk, and are flat in its outer regions.  Such a
shape of the rotation curve is typical for a large spiral galaxy. To build
more realistic models, we take into account the gravitational potentials
of the halo, of the bulge and of the galactic disk. The parameters of
the potentials and corresponding properties of the rotation curves are
listed in Table~\ref{tab::NbtMod}. To take into account observational uncertainties in our knowledge of the rotational curve of the galaxy, we vary the rotational velocity of the disk. Namely, the model A has the largest rotational velocity, B -- middle one and model C has the lowest rotational velocity. The resulting rotation curves are
shown in Fig.~\ref{fig::diff_rc}. A large spread in the rotational
velocities reflects an uncertainty of observational data for this galaxy.

\subsection{Velocity dispersion}
To constrain the radial component of the velocity dispersion of the
stellar disk $c_r$ we use observed  line-of-sight velocity dispersion data of
NGC 5247. For the nearly face-on
orientation of NGC 5247, the line-of-sight velocity dispersion is close
to the component of the velocity dispersion in the direction perpendicular
to the disk.

\begin{figure}
\includegraphics[width=1.0\hsize]{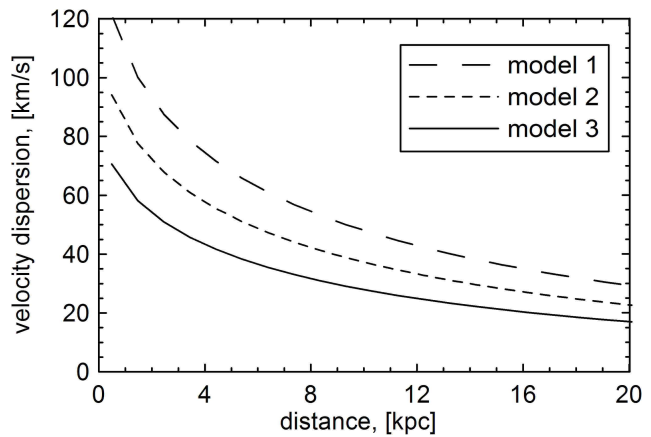}
\caption{Radial velocity dispersion as a function of galactocentric distance. In model $1$ ---
$c_z/c_r = 0.43$, model $2$
--- $c_z/c_r = 0.6$, model $3$ ---
$c_z/c_r=0.8$}\label{fig::diff_cr}
\end{figure}

To mimic the observational uncertainties in our knowledge of the velocity dispersion, we choose three profiles as shown in Fig.~\ref{fig::diff_cr}. In this Figure, index 1 corresponds to the hottest disk, and index 3 corresponds to a model with the lowest velocity dispersion.

Observational data of the external galaxies show that the ratio
of the vertical velocity dispersion to the radial velocity dispersion
is approximately constant along the radius of a galactic disk, and
varies from galaxy to galaxy within $c_z/c_r=0.3-0.8$  \citep{Korchagin2000,Khoperskov2010}. In our models we choose this ratio equal to $c_z/c_r = 0.43;0.6;0.8$ in central regions of galaxy.

\subsection{Disk thickness}

A direct measurement of the vertical scale of the disk of NGC 5247 is impossible. \citet{Zhao2006} solved this
problem by applying the Jeans equation in the direction perpendicular to
the collisionless stellar disk. They found the value of the disk
vertical scale to be $z_0 = 1.5 \pm 0.6$ kpc. Taking into account this
result, we additionally consider two variations of the model C3
with larger thickness of the disk equal to $z_0 = 2.1$ kpc,
and the lower disk scale length $z_0 = 0.9$ (model D3).

A variation of the thickness of the stellar disk is taken into account in a set of models D with thickness of the disk decreasing from model D1 to model D3.

\section{One-component models}

\subsection{Hydrodynamical model}

The simplest model to study the dynamics of an unstable
galactic disk is the hydrodynamical approximation. In this
approximation, a stellar-gaseous gravitating disk is represented by a
gravitating compressible fluid. There are some arguments justifying the
hydrodynamic approximation for a description of stellar disks:
\citet{Marochnik1966}, \citet{Hunter1979}, and \citet{Sygnet1987}
demonstrate that the behavior of perturbations in collisionless disks
can be approximately described by introducing an isotropic pressure with
polytropic constant $\gamma=2$. Using the hydrodynamic approximation
\citet{Kikuchi1997} made a direct comparison of the global stability
properties of gravitating disks with the solution of the collisionless
Boltzmann equation given by \citet{Vauterin1996}. \citet{Kikuchi1997}
found a qualitative, and in most cases a good quantitative agreement
between the results found in a hydrodynamic approximation and by the
direct solution of the Boltzmann equation.

To study the dynamics of the disk of NGC 5247, we use the parameters
of model C1, listed in Table \ref{tab::NbtMod}. This model was studied
by using the linear global stability analysis as well as two-dimensional nonlinear simulations.
Both methods are described in detail in papers
by~\citet{Korchagin2000} and \citet{Korchagin2005}. The linear analysis
yields the most unstable global mode as the $m=2$ mode with
eigenvalues $\rm Re(\omega_2)
= 1.67$, $\rm Im(\omega_2)=2.77$. Its nearest competitor, $m=3$ mode, has the
imaginary part of the eigenfrequency that determines the mode's growth rate
equal to  $\rm Im(\omega_3)=2.43$. With such values, the dynamics of perturbations
is determined by a two-armed spiral pattern.

To analyze quantitatively the large-scale morphology of the spiral
 structure formed in the disk, we use a Fourier decomposition of
 the surface density $\Sigma(r,\varphi) = \int_{-\infty}^{\infty}\varrho_s(r,\varphi,z)dz \, $ of the disk.

\begin{figure}
 \includegraphics[width=1.0\hsize]{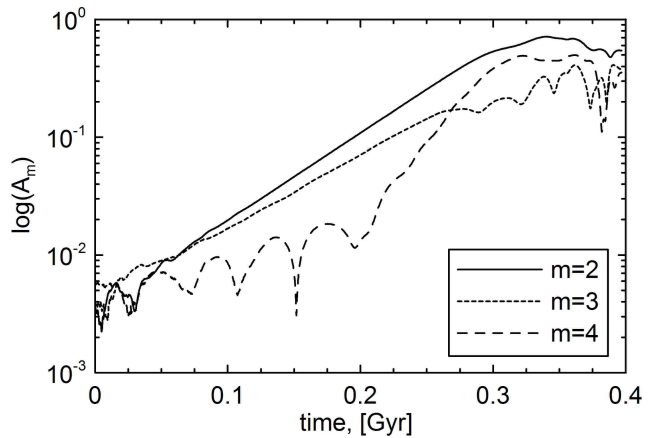}
 \caption{Time evolution of
 the $\log_{10} (A_m)$ for $m=2-4$ global Fourier amplitudes. The most
 unstable mode is the $m=2$ spiral. The instability saturates at level
 of $\log_{10} (A_m) = 0.5$.
 }
 \label{fig::FA}
\end{figure}

\begin{figure}
 \includegraphics[width=1.0\hsize]{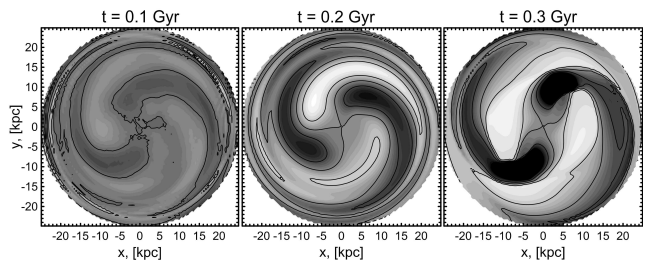}
 \caption{Time evolution of density perturbations in galactic disk in
 model C1. The spiral is in a counterclockwise rotation}
 \label{fig::H2}
\end{figure}

Figure \ref{fig::FA} plots the time dependence of the global Fourier
amplitudes for modes $m=1-4$
\begin{equation}\label{eq::Four_Cm}
A_m = \frac{\left| \int_0^{2\pi} \int_0^{1}
\Sigma(r,\varphi)e^{-im\varphi}r dr d\varphi
\right|}{\int_0^{2\pi} \int_0^{1} \Sigma(r,\varphi)r dr
d\varphi}\,
\end{equation}
growing from random initial density
perturbations.  Snapshots of density perturbations in the nonlinear
simulations are shown in Figure (\ref{fig::H2}), which clearly
illustrates how the two-armed spiral structure evolves out of the initial
random density perturbation.

\subsection{Collisionless models}

Although the hydrodynamical approximation reproduces qualitatively the
morphology of NGC 5247, the lifetime of the spiral pattern
is short. After approximately 0.3 Gyrs the spiral pattern is destroyed
in these hydrodynamical models and they no longer resemble
the observed spiral structure.
A mechanism that destroys the spiral structure is not completely understood.
Using hydrodynamical model, \citet{Laughlin1997} showed that a nonlinear spiral mode
causes a hollowing out of the surface density profile in a vicinity of a corotation
resonance. This stops the spiral growth and eventually destroys the spiral itself.
For the collisionless disks, the mechanism remains unclear.
 We explore therefore the collisionless
models to study to what extent the collisionless nature of the
galactic disk increases the lifetime of the spirals.

\begin{figure}
\includegraphics[width=1.0\hsize]{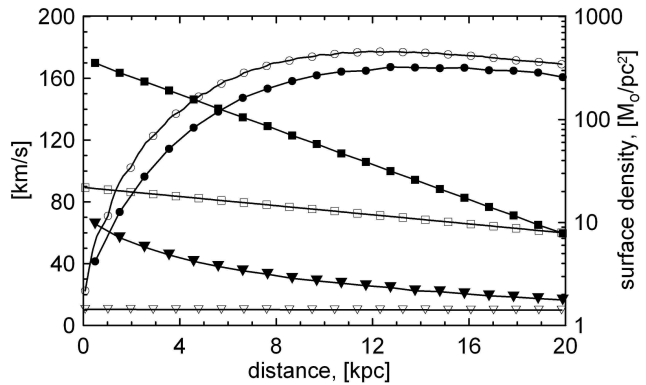}
\caption{Initial distributions of the stellar-gaseous disk
parameters in model E2: velocity of stars (open circles), velocity of gas (fill circles), radial velocity dispersion of stellar disk (fill triangles), speed of sound in gaseous component (open triangles), stellar surface density (fill squares) and gaseous surface density (open squares)}\label{fig::ini_st_gas}
\end{figure}

Using distributions shown in Figure \ref{fig::ini_st_gas} we build a set
of models for a stellar disk that have stability $Q$-parameter larger
then unity:

\begin{equation}\label{eq::ToomreSt}
Q_{Ts} = \frac{\kappa c_r }{3.36 G \Sigma} \geq 1\,.
\end{equation}

If condition (\ref{eq::ToomreSt}) does not hold for the whole disk,
then the perturbations grow significantly during a short evolution
time, and the parameters of the disk change within 1--2 disk
rotations. We exclude such cases from our consideration.

\begin{figure*}
\includegraphics[width=1.0\hsize]{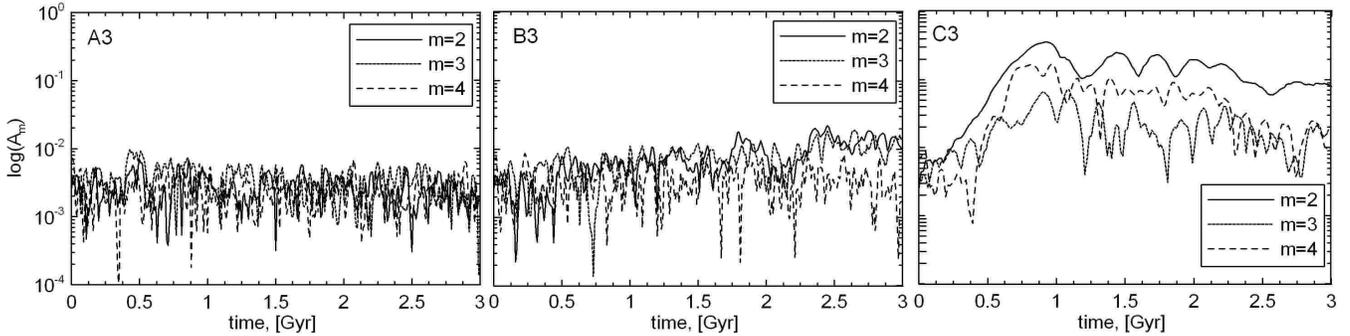}
\caption{The amplitudes of the Fourier harmonic for models with
different rotation curves. Left frame --- Model A3, central
frame --- Model B3 and right frame --- Model C3. Parameters of the models are given
in Table 1.}
\label{fig::FourABC}
\end{figure*}

The rotation curve significantly influences the dynamics of a
collisionless disk. Figure \ref{fig::FourABC} shows evolution of
the Fourier harmonics $A_m$ in three models with different rotation
curves. In the fast rotating model A3, the disk is stable due to
the presence of a massive halo. The model with the fastest disk rotation
therefore can not explain the observed spiral pattern in the galaxy NGC
5247. In model B3 (middle frame), the perturbations slowly grow with
the most unstable mode two-armed spiral.
During approximately three Gyr from the beginning of
simulations, the wave amplitude reaches about 2--5\% of disk unperturbed
values which is significantly lower then the amplitude of the observed
spiral pattern. Apparently, the models A and B can not explain the
observed spiral structure of the galaxy NGC 5247.

\begin{figure}
\includegraphics[width=1.0\hsize]{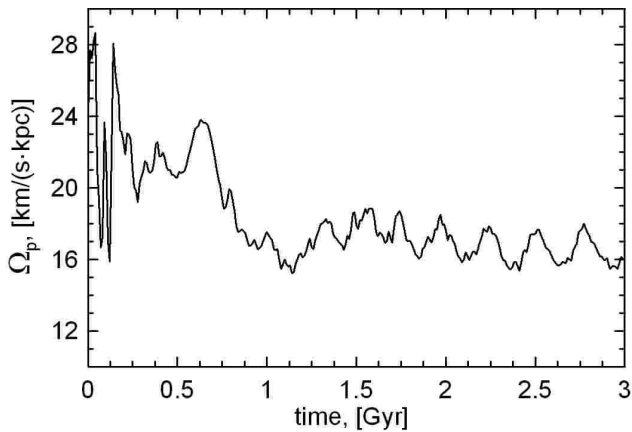}
\caption{Time dependence of angular velocity of two-armed pattern in model C3.}\label{fig::omegap2}
\end{figure}

Let us examine model C, which has a more massive stellar disk
compared to models A and B. The minimum value of the Toomre
$Q$-parameter indicates that this model is gravitationally
unstable. Similar to model B, there is a well defined two-arm
global spiral structure that extends to about four scalelengths in
the disk. The amplitude of the perturbations is much higher
compared to the other cases. During the initial stages
($t<0.5 \times 10^9$ years) one can see a
three-arm spiral, and the spiral pattern is a superposition of a
two- and three-armed modes. However, in the nonlinear stage a
two-armed pattern dominates.  Figure \ref{fig::omegap2} shows a typical dependence of an angular velocity of a two-armed spiral
pattern generated in model C3. Between approximately $0.2 - 0.7$ Gyr when an exponential growth of a spiral perturbation
is observed, a spiral perturbation has an angular velocity equal approximately to 22 km s$^{-1}$ kpc$^{-1}$. After 0.8 Gyr, a nonlinear
saturation starts, the angular velocity decreases to 18 km s$^{-1}$ kpc$^{-1}$ and remains approximately constant during the rest
of simulations.

\begin{figure}
\includegraphics[width=1.0\hsize]{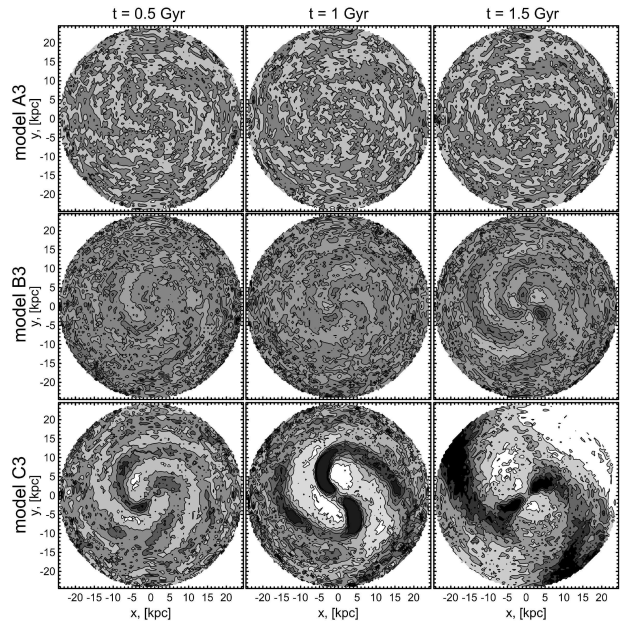}
\caption{Evolution of the surface density in the numerical simulations of the collisionless disks with different rotation curves}\label{fig::rc_evol}
\end{figure}

\begin{figure}
\includegraphics[width=1.0\hsize]{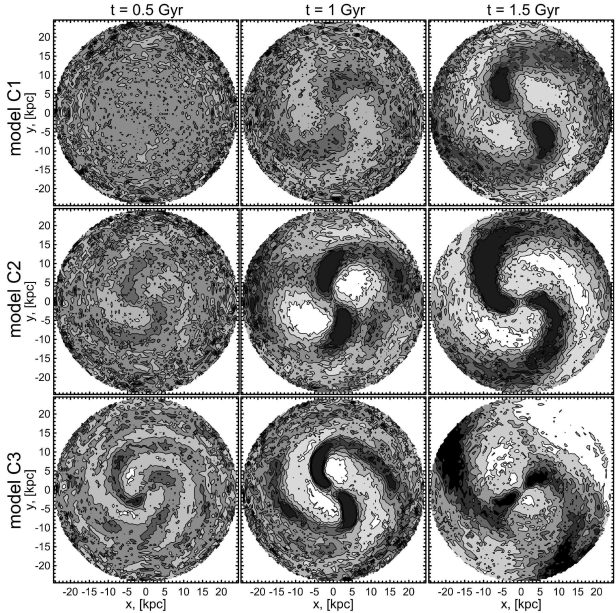}
\caption{Evolution of the perturbed surface density of the stellar disk in simulations with different radial velocity dispersion profiles.}\label{fig::cr_evol}
\end{figure}

\begin{figure*}
\includegraphics[width=1.0\hsize]{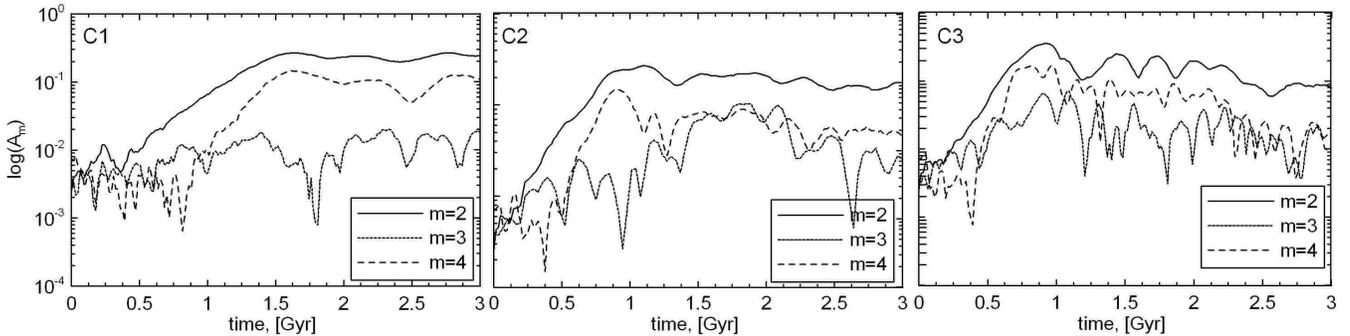}
\caption{Fourier harmonics in models with different radial velocity dispersions: left frame
$c_z/c_r=0.8$, central frame $c_z/c_r=0.6$, right frame
$c_z/c_r=0.43$.}\label{fig::FourC123}
\end{figure*}

Despite significant difference in the values of radial velocity
dispersion used in our numerical simulations, the spiral patterns
are quite similar as can be seen by comparison models C1, C2,
and C3 in Figure \ref{fig::cr_evol}.
On a qualitative level one can conclude that
morphological properties of spiral structure are independent on variation of the
radial velocity dispersion within observational errors. The
growth rates however are  different  ranging
from 0.7 Gyr in the coldest disk to 1.5 Gyr for model
with the largest velocity dispersion (Figure \ref{fig::FourC123}).

\begin{figure}
\includegraphics[width=1.0\hsize]{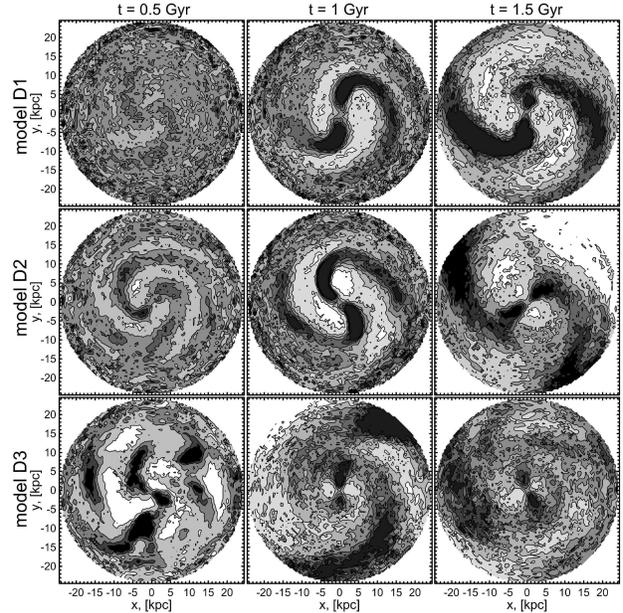}
\caption{Evolution of the surface density perturbations in stellar disk in numerical simulations with different thickness of the stellar disk: D3 -- $z_0=0.9$ kpc, D2 -- $z_0=1.5$ kpc, D1 -- $z_0=2.2$ kpc.}
\label{fig::widthevol}
\end{figure}

The value of the disk vertical scale used in our simulations
is close to the value estimated by \citet{Zhao2004} $1.1-
2.5$\,kpc, implying a large range of values to be explored.
Figure \ref{fig::widthevol} shows that the spiral patterns in
collisionless cases D1--D3 are similar, and independent
of the disk thickness.

Taking into account the third dimension, i.e. the disk
thickness, is important in disk dynamics to simulate correctly the density
growth in spiral arms. The vertical motions lead to a decrease of
the growth rate of an unstable spiral wave. Under similar
conditions, the amplitude of the spiral wave grows faster in a
thinner disk (Fig.~\ref{fig::widthevol}). In the model with a
relatively large
scale length of $z_0=2.1$\,kpc, the structure of the spiral arms and
their lifetimes are qualitatively similar to the behavior of
model C3. However, the disk with a small thickness is
significantly more unstable, resulting in a rapid growth of
multi-armed perturbations and formation of a flocculent
spiral pattern.

In general, the lifetime of the global two-armed spiral pattern in
our collisionless models is about 1.5 Gyr. As was mentioned,
a gaseous component in the disk increases the lifetime of the spiral
structure. In the following section we will consider this effect in detail.

\section{Two-component models}

In our two-component models, we take into account the gravitational
interaction between the gaseous and the stellar components of the disk.
Effects of star-formation or exchange of matter between
the stellar and gaseous subsystems are neglected.
It is known that gas has a strong
destabilizing effect on a gravitating disk even if its surface density
is much smaller than the surface density of the stellar disk.
We build our
stellar-gaseous models based on the one-component model C3, and
by varying the
total gas mass within the radius of the stellar disk. Namely, we analyze
the models with  the relative mass of a gaseous component equal to:
$$
M_g/M_s = \left\{ 0.01; 0.05; 0.1; 0.2 \right\}\,.
$$

For a two-component gravitating disk, each subsystem has its own
spatial density distribution, rotation curve and velocity dispersion
profile. Figure \ref{fig::ini_st_gas} shows initial equilibrium
distributions of basic quantities for a stellar-gaseous disk. A difference
between the rotation curves of stellar and gaseous components is caused
by a difference in their radial velocity dispersions, and influence of the
velocity dispersions on disk equilibrium. The stellar disk has
an exponential
density distribution with a radial scale length of $r_d=4.8$ kpc.
The velocity dispersion of the stellar disk in the direction
perpendicular to the disk decreases from 70 \kmps\ in its central regions to
about 20
\kmps\ at disk periphery. The gaseous disk is cold compared to
the stellar one. Its velocity dispersion is close to 10 \kmps\
in the disk's central regions and changes slowly within the stellar disk.

\begin{figure}
\includegraphics[width=1.0\hsize]{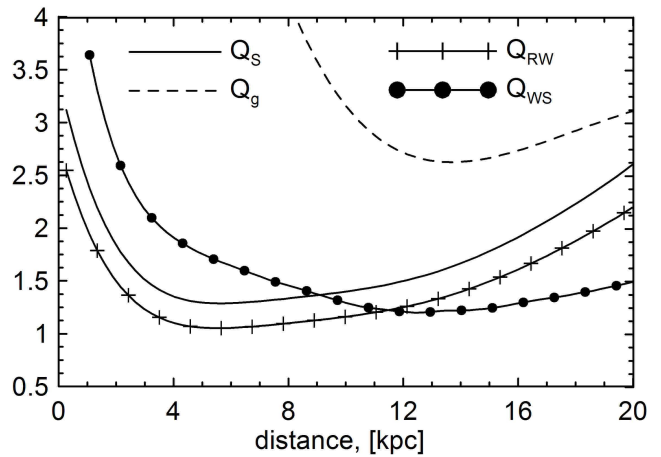}
\caption{Radial distribution of Toomre parameter: gaseous, stellar and both stellar-gaseous disk for model E2}.\label{fig::Toomre_st_gas}
\end{figure}

For a two component gravitating disk, the stability criterion needs to
be modified. A few such modifications of the stability parameter have been
suggested in the literature. A simple stability criterion has been
proposed by \citet{WangSilk1994}:

\begin{equation}\label{eq::Toomre_WS}
\frac{1}{Q_{WS}} = \frac{1}{Q_{s}} + \frac{1}{Q_{g}}\,.
\end{equation}
Here $Q_s$ and $Q_g$ are the values of the Toomre $Q$-parameter for the
stellar and the gaseous components respectively. \citet{Romeo2011}
demonstrated that such an approach is not suitable for
three-dimensional systems. \citet{Romeo2011} suggested  an
expression for the effective stability
criterion that takes into account the finite thicknesses of a stellar and
a gaseous disk:
\begin{equation}\label{eq::Toomre_RW}
 \frac{1}{Q_{RW}} = \left\{%
\begin{array}{ll}
    \displaystyle \frac{W}{T_s Q_{s}} + \frac{1}{T_g Q_{g}}\,, & { T_s Q_s \geq T_g Q_g;} \\
    \displaystyle \frac{1}{T_s Q_{s}} + \frac{W}{T_g Q_{g}}\,, & { T_g Q_g \geq T_s Q_s\,.} \\
\end{array}%
\right.
\end{equation}
Here $ W = {2 c_r c_s}/{(c_s^2 + c_r^2)}$, $ T_s =
0.7+0.8({c_z}/{c_r})_s$, $ T_g=0.7 + \left({c_z}/{c_r}\right)_g$.
The radial dependence of the stability parameters $Q_s$, $Q_g$ and
$Q_{RW}$ are shown in Figure \ref{fig::Toomre_st_gas}.

\begin{figure}
\includegraphics[width=1.0\hsize]{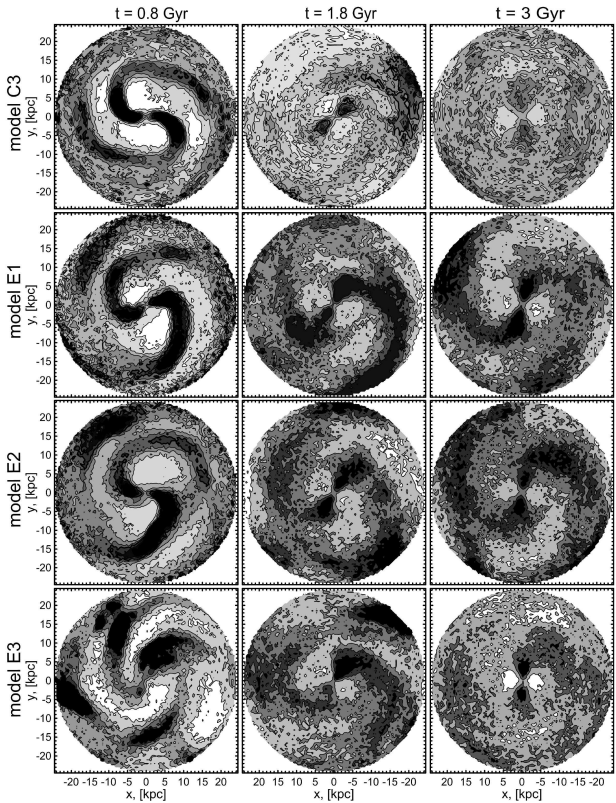}
\caption{Comparison of the evolution of the surface density perturbations of a stellar disk in numerical simulations with different amount of a gas-component}\label{fig::gas_evol}
\end{figure}

The evolution of the surface density perturbations in a stellar
disk in models with different admixtures of gas is shown in
Fig.~\ref{fig::gas_evol}.  During $t<1$ Gyr, a spiral pattern is formed
in all models. Later on ($t>1.5$ Gyr), the spiral wave dissipates in the
purely collisionless model, and becomes indistinguishable from
a background of low-scale density perturbations. A small gas admixture
with a mass of a few percent of the total mass of the disk
significantly increases the life time of the spiral structure, and
spiral pattern exists up to $~3$ Gyr.  If mass of the gas is
increased to one-tenth of the total mass of the disk, then a strong
gravitational instability is created, leading to the formation of
a complex, highly transient multi-armed spiral structure. The disk changes significantly from its initial state during about $0.5$ Gyr. In such
a case, star-formation should be taken into account which is beyond the
scope of our paper.

The self-consistent stellar-gaseous models allow reproducing a
long-living two-armed spiral pattern. We find however, similar
to \citet{Elmegreen1993} and \citet{Thomasson1990}, that the morphology of
the spiral pattern evolves with time, with overall lifetime of the
spirals being about ten disk rotations. In comparison, lifetime of
spirals in the one-component models do not exceed $1.5$ Gyr.

\begin{figure}
  \includegraphics[width=1.0\hsize]{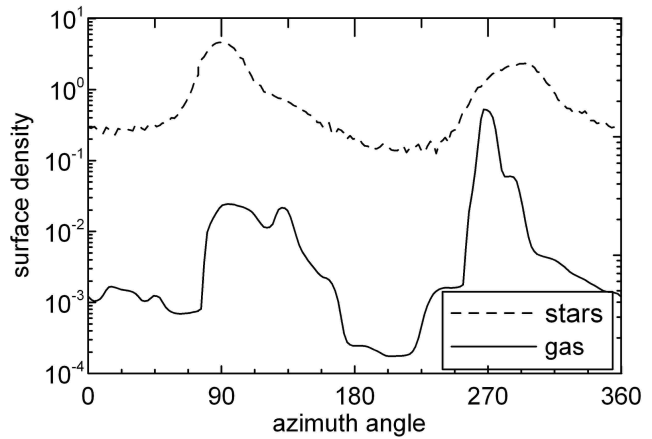}\caption{Azimuthal layer of the gaseous and stellar surface density at $t=1.5$ Gyr and $r=4.9$ kpc for model E2}\label{fig::shock_waves}
\end{figure}

Figure~\ref{fig::shock_waves} shows the azimuthal density
distribution at $r=4.9$ kpc in the gaseous, and in the stellar components.
The figure clearly demonstrates formation of shock fronts in the gaseous
component located at the inner edge of the stellar arms.

The shock front is formed when the supersonic flow of
interstellar gas passes through the density wave of the spiral arm
\citep{Roberts1969}. Distribution of gas along the azimuth indicates
the position of the shock wave on the inner edge of the stellar density
wave in our simulations. In the presence of a strong shear flow the
system of shock fronts is unstable to hydrodynamical instabilities
\citep{Wada2004,Khoperskov2011}. We can not reveal such an instability
due to the low resolution of our simulations.

\begin{figure}
 \includegraphics[width=1.0\hsize]{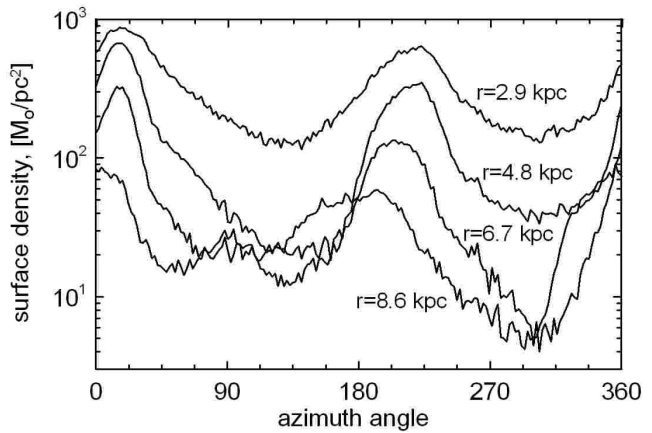}\caption{Azimuthal layers of a stellar-component surface density for various values of radius --- from top to bottom: $r= \{2.88; 4.8; 6.72; 8.64\}$ kpc for model E2.}\label{fig::1d_waves}
\end{figure}

Figure \ref{fig::1d_waves} shows azimuthal dependence of the stellar surface density at different radii in the disk. As one can judge from the figure, the
amplitude of the spiral perturbation decreases with radius in a stellar
component. The shift in phase of the density maxima with radius
reflects the spiral nature of the perturbations.

\section{Discussion}
\subsection{Toomre's parameter}
\citet{Sellwood2011} reviewed observational as well as theoretical
arguments and came to the conclusion that spiral patterns are
short-lived. A qualitative counterargument  against this picture
is that if spirals recur often enough a large fraction of galaxies
should have a low-amplitude spirals. This is not the case in the most of observed cases.
Below we discuss in detail two factors, considered by Sellwood
as arguments against long-lived spiral structure.

One of his arguments is that the theory of spiral modes requires
galactic disks to be dynamically cool. Namely, Toomre's axisymmetric
stability parameter should range within $1\le Q \le 1.2$. As a test of
applicability of a modal approach in such systems, Sellwood simulated
the dynamics of a disk that has Q-parameter equal to unity everywhere
except the disk's central regions where the Q-parameter rises steeply.
Via this experiment Sellwood demonstrated that such a low value
of Q-parameter supports
vigorous collective responses that change the disk properties during a
short dynamical time.

\begin{figure}
  \includegraphics[width=1.0\hsize]{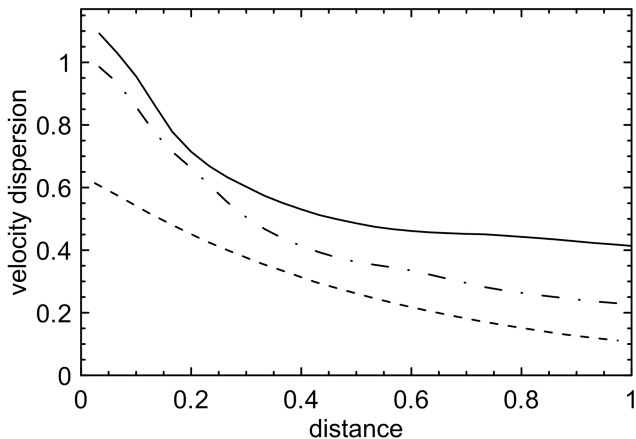}\caption{Radial distribution of the radial velocity dispersion at different times: dotted line is initial statement ($t=0$), dash-dot line is distribution after one rotation period ($t=T$), full line is stationary distribution ($t/T= 3\div\,10$).}\label{fig::crtcold_new}
\end{figure}

We repeat the experiment of Sellwood in 3D simulations assuming that the disk
of the galaxy NGC 5247 has $Q=1$ everywhere except its central regions. We
find a qualitative agreement of our simulations with the results of
\citet{Sellwood2011}. During approximately one disk rotation a complex
multi-armed transient spiral structure is formed that causes disk heating
and a quick dissipation of the spiral waves. Figure \ref{fig::crtcold_new}
shows the radial profile of disk radial velocity dispersion at different moments
of time. After a violent disk heating, it comes to a stationary stage.

The assumption about a radial profile of Toomre's $Q$-parameter invoked
by Sellwood is not related to the spiral galaxies. It is known from the
direct measurements that velocity dispersion in the galactic disks decreases exponentially with
radius and has a larger scale length compared to the scale length of the
surface brightness (or density). In combination with the fact that the
disks are nearly flat and self-gravitating in the vertical direction,
this leads to a bell-like shape of Toomre's $Q$-parameter rising towards
the center of a disk and on its periphery.  With such a $Q$-profile, the
disks are globally unstable and support a spiral structure during tens
of galactic rotations.

We note also that if real galaxies were to have a $Q$-parameter profile close
to that assumed by Sellwood, they would be in a phase of violent
evolution and have a very short dynamical time. This would be
be difficult to reconsile with the number of observed
spiral galaxies and their lifetimes.  The density wave theory (the
disk's instability towards the global spiral modes) does not require
the $Q$-parameter to be in a narrow range $1 \le Q \le 1.2$ as it was
demonstrated by a number of examples in hydrodynamical simulations
by \citet{Korchagin2000,Korchagin2005} and in collisionless models by
\citet{Khoperskov2007}.

\subsection{Resonances}

After the work of \citet{Toomre1981}, the swing amplification
of perturbations in a differentially rotating gravitating disk is
regarded as one of the key mechanisms that amplifies the spirals.
The swing amplification works in the presence
of a shearing disk. The orbital clock is faster at the interiors
of the galactic disks, and the mechanism runs more rapidly
at smaller radii. The important part of the swing amplification
mechanism is a propagation of the waves in a radial direction with
a group velocity. A propagating wave reflects from the center and/or
from a corotation region, changing from trailing to leading, and
simultaneously changing a sign of the group velocity and serves,
according to Toomre, as ``fresh grist to a swing amplified mill''
to operate continuously in a disk.

\begin{figure*}
\includegraphics[width=0.5\hsize]{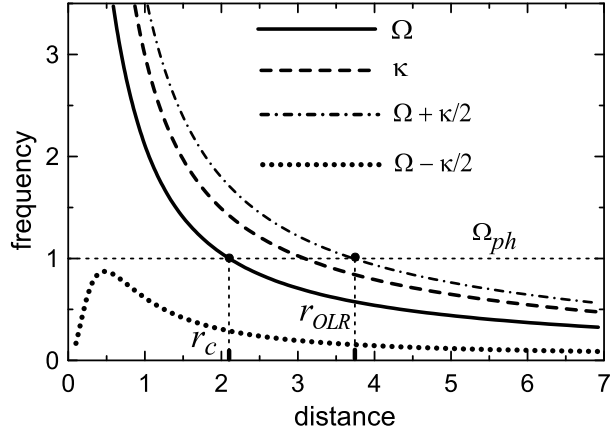}
 \caption{
Angular velocity, epicyclic frequency, and the positions of principal
resonances for a gaseous disk~($r_c$ -- corotation radius, $r_{olr}$ is the outer Lindblad resonance). The inner Lindblad resonance is not
achieved in this model. The units of length and frequency are
2 kpc and 72 km s$^{-1}$ kpc$^{-1}$ correspondingly.}
\label{fig::omegacor}
\end{figure*}

\begin{figure*}
\includegraphics[height=0.5\hsize]{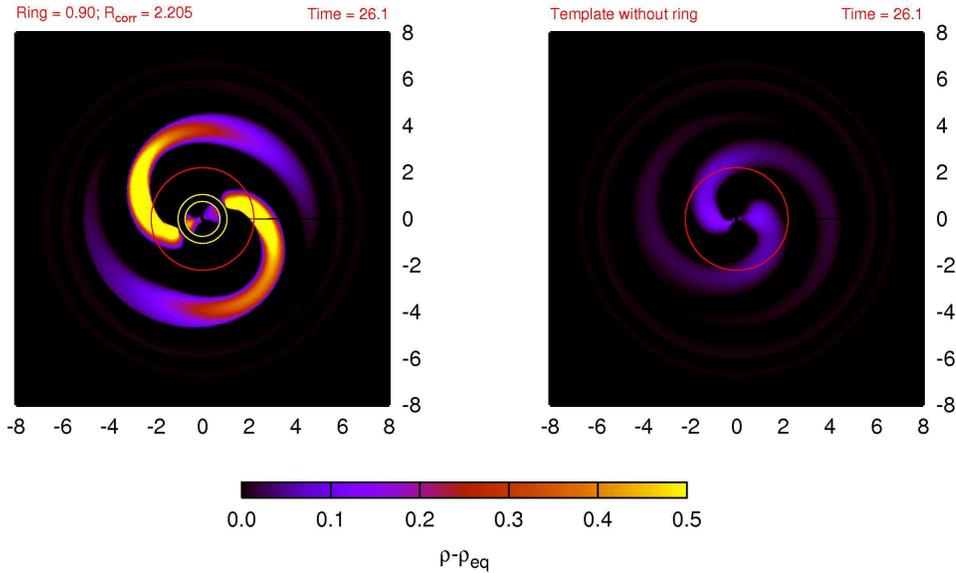}
 \caption{
 The right panel shows a two-armed spiral pattern
 developing in an unstable "template" gravitating disk (see section 7.1
 for details). The location of the
 corotation radius is shown by red circle in both the right and the left
 panels. The  left panel shows the results of a simulation when an
 absorption ring (whose inner and outer extent are shown by yellow circles)
 has been imposed close to the disk center. Note that the
 development of the 2-armed spiral pattern is not inhibited by the presence
 of the absorption ring, in contradiction to the expectations from standard
 swing amplification mechanism.}
\label{fig::abs_ring}
\end{figure*}

In this mechanism, the inner Lindblad resonance is considered as
the most important in generation of the spiral density waves. This
resonance must be shielded to prevent a fierce damping of the waves.
To check the importance of the inner Lindblad
resonance in generation of spirals, we performed an artificial
experiment with our hydrodynamical model  imposing an absorption
region in the inner regions of the disk where all perturbations are
zeroed out.This cuts the feedback loop and destroys
amplification of spiral perturbations.  We performed a series of
2D simulations of the unstable 2-D gaseous disk with such an
artificially imposed "absorption ring" where all perturbations are
forced to be zero and macroscopic characteristics of the disk are
set to be unperturbed in a smooth way. Fig.~\ref{fig::omegacor} shows
angular velocity, epicyclic frequency and positions of resonances for such a disk.
As one can see, the model does not harbor the inner Lindblad resonance,
so an absorption ring was imposed artificially to mimic its influence.

 Fig.~\ref{fig::abs_ring}
shows a result of such simulations.  On the right panel of
Fig.~\ref{fig::abs_ring} is shown a two-armed spiral pattern
developing in an unstable 'template" gravitating disk. The  left
panel shows the results of simulations when an absorption ring has
been imposed close to the disk center.  The disk remains unstable
and a two-armed spiral is developing out of noise perturbations.
We find that the spiral perturbations develop in the disk independently
of the position of the absorption ring, be it inside, or outside the corotation
resonance.

\begin{figure*}
\includegraphics[height=0.5\hsize]{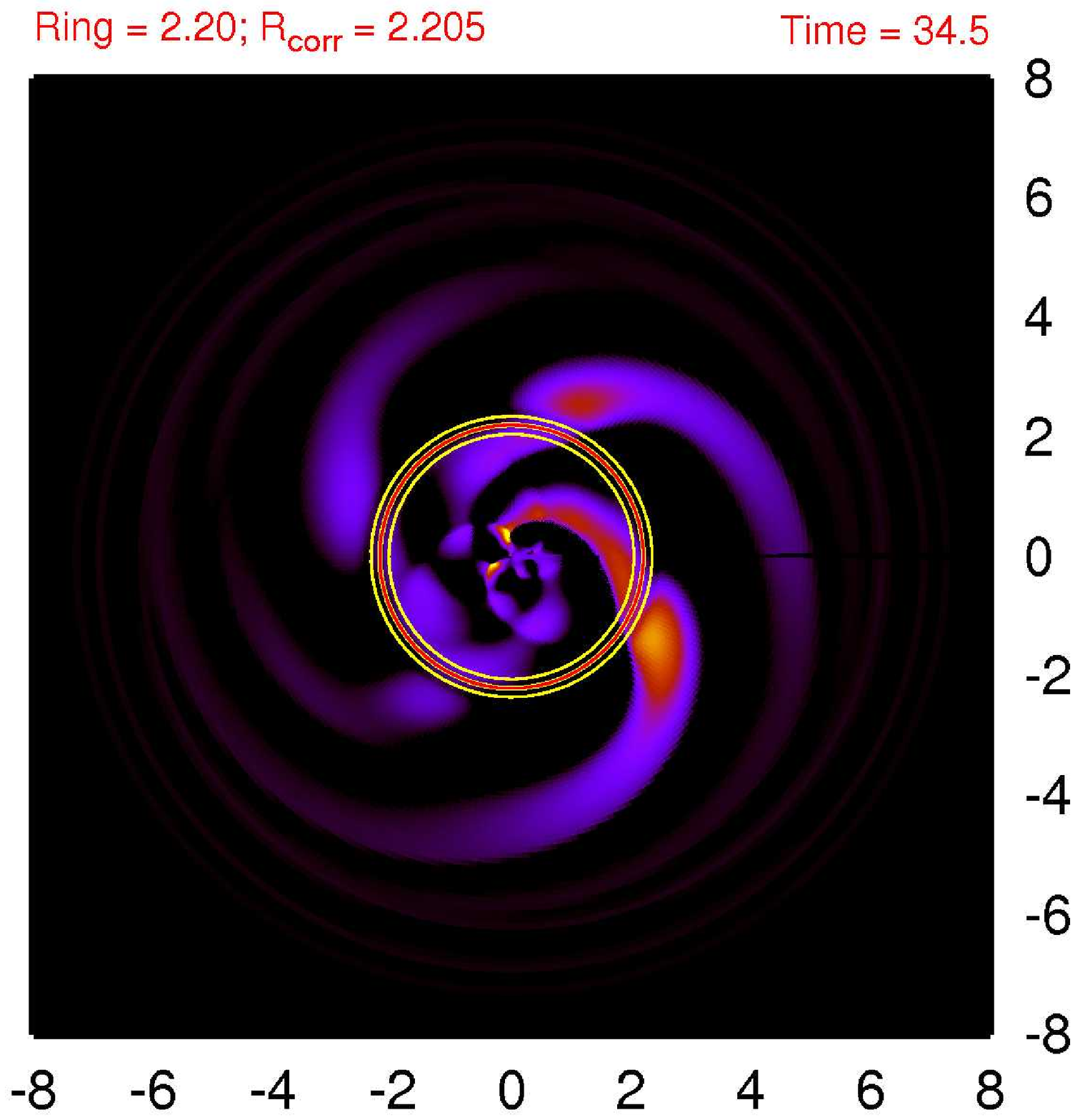}
 \caption{
 The figure shows the results of a simulation when an
 absorption ring (whose inner and outer extent are shown by yellow circles)
 has been imposed at the corotation radius. Note that unlike
 Fig.~\ref{fig::abs_ring}, growth of a two-armed spiral is completely
 suppressed due to the presence of the absorption ring, and a multi-armed
 spiral starts to form.
 }
\label{fig::cmp}
\end{figure*}

An instructive experiment is to impose the absorption ring at corotation.  The behavior of perturbations
changes dramatically in this case (Fig.~\ref{fig::cmp}).
The growth of a two-armed spiral is totally suppressed, and a
multi-armed spiral starts to grow with a four-armed spiral prevailing.
These experiments illustrate that the qualitative picture of spiral growth by swing amplification is not necessarily as relevant as once thought.
Instead, the corotation resonance may be more important in spiral generation.

The issue of transient spirals is currently a topic of heated debate. \citet{Zhang1996,Zhang1998,Zhang1999} finds that galaxies undergo a secular evolution
that leads to a redistribution of the disk matter and heats the disk stars so they gradually rize above the galactic plane in the disk central
regions. Both effects are the natural ingredients of a presence of the spiral density wave. Stars migrate inside and outside the corotation
region, and a more and more centrally concentrated density distribution is achieved with time together with the buildup of of an extended
outer envelope. Similar effect was found in hydrodynamical simulations by \citet{Laughlin1997}.

\section{Conclusion}

We have built one-component collisionless and hydrodynamical
models as well as multi-component stellar-gaseous models of the
disk galaxy NGC 5247.  The models are unstable to $m$=2 global
modes that form a two-armed spiral pattern with amplitudes of about
10--20\%, qualitatively similar to the observations.

Our simulations show that in purely collisionless models
the lifetime of the spiral structure is about a few galactic
rotations, and does not last longer than one Gyr. An admixture of a gaseous component with mass of a few
percent of the mass of the stellar component significantly increases
the lifetime of the spiral structure. In the simulated model
of galaxy NGC 5247, the spiral lifetime increases up to 3
Gyr, and comparable with the lifetime of galaxies.

\section{Acknowledgements}
The authors would like to thank an anonymous referee for many useful comments and questions.
The authors also express  their gratitude for financial support (~grants RFBR 11-02-12247-ofi-m-2011, 10-02-00231, 12-02-00685-a, 12-02-31452) and the Program for State Support for Leading Scientific Schools of the Russian Federation (grant NSh-3602.2012.2).
Numerical simulations were run at the Scientific Research Computing Center on supercomputers "Lomonosov" and "Chebyshev". S.~Khoperskov expresses his gratitude to non-commercial foundation of Dmitry Zimin "Dynasty" for financial support.

\begin{table*}
 \centering
 \begin{minipage}{140mm}
 \caption{The parameters of one-component (N-body) and two-component models. $c_{r0}$
is the radial velocity dispersion in center of the disk, $V_{max}$
--- the maximum value of the rotation velocity, percentage of gas - $M_g/M_s$, one-component model if equal zero.}\label{tab::NbtMod}
\begin{tabular}{|c|c|c|c|c|c|}
\hline
Model name & $c_z/c_r$ & $c_{r0}$ (km/s) & $z_0$ (kpc) & $V_{max}$ (km/s)& $M_g/M_s$\\
\hline
A1 &  0.43 & 120 & 1.5 & 245 & 0   \\
A2 &  0.6  & 92  & 1.5 & 245 & 0   \\
A3 &  0.8  & 70  & 1.5 & 245 & 0   \\
B1 &  0.43 & 120 & 1.5 & 200 & 0   \\
B2 &  0.6  & 92  & 1.5 & 200 & 0   \\
B3 &  0.8  & 70  & 1.5 & 200 & 0   \\
C1 &  0.43 & 120 & 1.5 & 153 & 0   \\
C2 &  0.6  & 92  & 1.5 & 153 & 0   \\
C3 &  0.8  & 70  & 1.5 & 153 & 0   \\
D1 &  0.8  & 70  & 2.2  & 153 & 0   \\
D2 &  0.8  & 70  & 1.5 & 153 & 0   \\
D3 &  0.8  & 70  & 0.9 & 153 & 0   \\
E1 &  0.8  & 70  & 1.5 & 153 & 0.01\\
E2 &  0.8  & 70  & 1.5 & 153 & 0.05\\
E3 &  0.8  & 70  & 1.5 & 153 & 0.1 \\
E4 &  0.8  & 70  & 1.5 & 153 & 0.2 \\
\hline
\end{tabular}
\end{minipage}
\end{table*}

\end{document}